\documentclass[journal=jacsat,manuscript=article]{achemso}

\usepackage[version=3]{mhchem} 
\usepackage[T1]{fontenc}       

\author{Kai Jiang}
\email{kaijiang@xtu.edu.cn}
 \affiliation[]{
 Hunan Key Laboratory for Computation and
 Simulation in Science and Engineering,
School of Mathematics and Computational Science
 (Xiangtan University),
 Hunan 411105, China}
\author{Juan Zhang}
\email{zhangjuan@xtu.edu.cn}
\author{Qin Liang}

\title [An \textsf{achemso} demo]
{Self-Assembly of Asymmetrically Interacting ABC Star Triblock
Copolymer Melts}

\abbreviations{IR,NMR,UV}
\keywords{American Chemical Society, \LaTeX}

\begin{document}


\begin{abstract}
The phase behavior of asymmetrically interacting ABC star triblock copolymer
melts is investigated by the self-consistent field theory (SCFT).
Motivated by the experimental systems, in this study,
we focus on the systems in which the
Flory-Huggins interaction parameters satisfy
$\chi_{AC}>\chi_{BC}\approx \chi_{AB}$.
Using various initialization strategies, a large number of
periodic structures have been obtained in our calculations.
A fourth-order pseudospectral algorithm combined with Anderson
mixing method is used to compute the free energy of candidate
structures carefully.
The stability has been detailedly analyzed by
splitting the free energy into internal and entropic parts.
A complete and complex triangular phase diagram is presented
for a model with $\chi_{AC}>\chi_{BC}= \chi_{AB}$ in which
fifteen ordered phases, including two-, and three-dimensional structures,
have been predicted to be stable from the SCFT calculations.
Generally speaking, with the asymmetrical interactions,
the hierarchical structures tend to be formed near the B-rich
corner of the triangular phase diagram.
This work broadens the previous theoretical
results from equal interaction systems to unequal interaction
systems.
The predicted phase behavior is in good agreement with experimental
observations and previous theoretical results.
\end{abstract}

\section{Introduction}
\label{sec:intr}

Block copolymers, constructed by linking together chemically
distinct subchains or blocks, spontaneously assemble into
exquisitely ordered soft materials\,\cite{bates2012multiblock, hamley2004developments}.
The self-assembled order structures, spanning length scales from a few
nanometers to several micrometers, offer a diverse and expanding
range of practical applications in, for example,
optical materials, microelectronic materials, drug delivery,
advanced plastics, and nanotemplates\,\cite{park2003enabling, meng2009stimuli,
ruiz2008density}.

The development of nanotechnology using block
copolymers requires a good understanding of the phase behavior of
the block copolymers.
The self-assembling mechanism of block copolymers sensitively depends on block types, the number of blocks,
block-block interactions, architecture, and topology of the block polymers.
The equilibrium ordered patterns can be formed due to
the delicate balance between these competing factors.
In ABC triblock copolymers, the number of controlled parameters is at
least five, including three interaction parameters
$\chi_{AB}N$, $\chi_{AC}N$, $\chi_{BC}N$, and
two independent block compositions $f_A$ and $f_B$.
$N$ is the degree of polymerization and $\chi_{\alpha\beta}$ is the Flory-Huggins interaction parameter
characterizing the interaction between two chemically different blocks $\alpha$ and $\beta$.
Compared with linear copolymers, the star-shaped copolymers have
complicated phase behavior or physical properties
induced by different molecular architecture.
In the ordered phases of ABC star triblock copolymers, the most distinct feature is
the arrangement of the junction points.
If the chain lengths of three blocks are comparable,
junction points are aligned on a one-dimensional (1D) straight line,
then cylindrical morphologies can be formed naturally.
Their cross sections tend to show two-dimensional (2D) patterns since
polymer/polymer interfaces can be flat surfaces due to the repulsion forces between ``unconnected'' branches.
These factors lead to the formation of 2D polygonal tiling patterns, or Archimedean tiling.
The tiling patterns can be encoded by a set of integers $[k, l, m, \dots]$, indicating
that a $k$-gon, an $l$-gon, and an $m$-gon, etc., meet
consecutively at each vertex.
If some asymmetry, to the contrary, is introduced to the
compositions, the junction points are mostly aligned on the curved
trails. Consequently three-dimensional (3D) structures can be formed.
Furthermore, when the interactions are strong
enough, the ABC star triblock copolymers can self-assemble into
hierarchical structures\,\cite{matsushita2010jewelry, matsushita2011kaleidoscopic}.

Owing to the rich phase behavior, a number of experiments
have been carried out on the phase and phase transition of ABC star triblock copolymers in past
decades\,\cite{okamoto1997morphology, sioula1998directrohtua,
sioula1998novel, takano2004observation, takano2005mesoscopic,
hayashid2006archimedean, matsushita2007creation, hayashida2007hierarchical,
matsushita2010jewelry, matsushita2011kaleidoscopic,
huckstadt2000synthesis, nunns2013synthesis, park2015morphology}.
In particular, Matsushita and co-workers have conducted systematic studies on the morphologies
formation of (polyisoprene-polystyrene-poly(2-vinylpyridine))
(ISP) star triblock copolymers. Several ordered tiling patterns, such
as [6.6.6], [8.8.4], [12.6.4], [3.3.4.3.4], and even a dodecagonal
symmetric quasicrystalline tiling\,\cite{hayashida2007polymeric}, have been observed in the ISP
star triblock copolymers\,\cite{takano2004observation, takano2005mesoscopic,
hayashid2006archimedean, matsushita2007creation, matsushita2010jewelry,
matsushita2011kaleidoscopic}. Meanwhile, several
hierarchical structures including cylinders-in-lamella,
lamellae-in-cylinder, lamellae-in-sphere, and
hierarchical double gyroid structures are also discovered with the
asymmetries of composition\,\cite{hayashida2007hierarchical,
matsushita2010jewelry, matsushita2011kaleidoscopic}.
Recently, Nunns et al. used polyisoprene (I),
polystyrene (S) and poly-(ferrocenylethylmethylsilane) (F) to
synthesize ISF star triblock
copolymers\,\cite{nunns2013synthesis}. Three 2D patterns,
including [8.8.4], [12.6.4], and lamellae with alternating
cylinders, were observed.
Both of the experimental systems satisfy asymmetrical interaction
between different blocks,
i.e., $\chi_{AC}N>\chi_{BC}N\approx \chi_{AB}N$.
In particular, in ISP star triblock copolymers, the interaction
strengths are known to follow the order $\chi_{IS}
N\approx\chi_{SP} N < \chi_{IP} N$\,\cite{matsushita2010jewelry},
while in ISF star-shaped triblock copolymers, $\chi_{SF}
N\approx\chi_{IS} N < \chi_{IF}N$\,\cite{nunns2013synthesis}.

Besides experimental works, theoretical studies provide a good
understanding of phase behavior of ABC star triblock copolymers.
Bohbot-Raviv and Wang\,\cite{bohbot2000discovering} used a
coarse-grained free energy functional to numerically investigate
some morphologies of ABC star triblock copolymers.
In 2002, Gemma and co-workers\,\cite{gemma2002monte} carried out Monte Carlo (MC) simulations on
ABC star triblock copolymers with equal interactions between the
three components. The phase behavior of ABC star triblock
copolymers with composition ratio of, $f_A:f_B:f_C = 1.0:1.0:x$, was investigated in
detail in strong segregation region.
Five kinds of 2D cylindrical phases, three kinds
of lamellar-type phases and two kinds of continuous matrix phases were obtained from the MC simulations.
Huang and co-workers\,\cite{huang2008morphological} studied the
effects of composition and interaction parameter on the phase
behavior of ABC star copolymers with equal interactions among the
three components using dissipative particle dynamics (DPD) simulations.
Several efforts have also been made to study the phase behavior
of ABC star triblock copolymers using the
SCFT\,\cite{tang2004morphology, zhang2010scft, li2010real,
xu2010stability, xu2013strategy}.
In 2004, Tang et al.\,\cite{tang2004morphology} started to use a 2D
SCFT simulation to study the phase behavior of ABC star triblock copolymers.
Based on the SCFT, Zhang et al.\,\cite{zhang2010scft} and Li et al.\,\cite{li2010real}
examined the weak and intermediate segregation cases of mainly 2D structures with equal interactions.
Zhang et al.\,\cite{zhang2010scft} also chose $\chi_{AB} N = \chi_{BC} N = 25.0$, $\chi_{AC} N =37.0$ to model
the ISP star triblock copolymer system of the type
$\mathrm{A_{1.0}B_{1.0}C_x}$. Accordingly, a 1D phase diagram was obtained as a
function of $x$ from $0.5$ to $2.0$.
The stability of the different lamellar morphologies formed from
ABC star triblock copolymers has been examined by Xu et al.\,\cite{xu2010stability}.
These SCFT studies mainly focus on the symmetrically interacting systems and
the phase behavior of 2D structures.
Despite these previous experimental and theoretical studies, a
comprehensive understanding of ABC star triblock copolymers is
still lacking, especially for the asymmetric interaction systems.
In this work, we will explore the phase behavior of ABC star triblock copolymers, including 2D and 3D
structures, with asymmetric interaction parameters between chemically different blocks.

Theoretical approaches to investigating the phase behavior of
block copolymers, often involve minimizing an appropriate free
energy functional of the system, and comparing the free energies
of different candidate structures.
Therefore a systematic examination of the emergence and stability of
ordered phases requires the availability of suitable free energy
functionals and accurate methods to compute the free energy of ordered phases.
Owing to a large number of studies, it has been well proven that the SCFT provides a powerful
theoretical framework for the study of the phase behavior of
block copolymers\,\cite{matsen2002standard,
fredrickson2006equilibrium}. In particular, the SCFT can be used to
determine the relative stability of different phases because it
provides an accurate estimate of the free energy. The essence of
the SCFT is that the free energy of the system can be
written as a functional of the spatially varying polymer
densities and a set of conjugate fields.
Minimizing the free energy functional with respect to the
densities and conjugate fields leads to a set of equations, encoded as SCFT equations.
The SCFT equations are a set of highly nonlinear equations with multi-solutions which correspond to the different ordered phases of block copolymers.
The equations also have a strong nonlocality that
emerges from the connection of propagators and densities, conjugate fields.

Solving the SCFT equations requires iterative techniques.
Owing to the nature of iterative methods, the solutions sensitively
depend on the initial configuration at the start of the iteration.
A series of efficient strategies of screening initial conditions are
developed based on the fact that all periodic structures belong
to one of $230$ space groups\,\cite{jiang2010spectral,
jiang2013discovery, xu2013strategy}.
In each iterative step, efficient numerical schemes are required to
solve the propagator equations.
In the past years, two complementary methods, including spectral methods\,\cite{matsen1994stable,
guo2008discovering} and real-space methods\,\cite{drolet1999combinatorial} to solve the SCFT equations have been developed.
In recent years, an efficient pseudospectral method has been introduced to
solve propagator equations\,\cite{rasmussen2002improved,
cochran2006stability, ranjan2008linear}.
This algorithm takes advantage of the best features of
real- and Fourier-space with the computational
effort scale of $O(N\log N)$ based on the fast Fourier
transformation (FFT). $N$ is the number of spectral modes or
discrete points in real-space.
To obtain the solutions of the SCFT equations corresponding to
the saddle points of the SCFT free energy functional, the
iterative methods are required to make the iteration convergent.
The quasi-Newton methods were employed in the fully spectral approach
to the SCFT by Matsen and Schick\,\cite{matsen1994stable}.
A simple mixing method by a linear combination of two
consecutive fields was introduced by Drolet and Fredrickson\,\cite{drolet1999combinatorial} for the SCFT simulations.
Recently the Anderson mixing method\,\cite{schmid1995quantitative, thompson2004improved} has been proven by itself to greatly reduce the number of SCFT iterations.
From the perspective of nonlinear optimization,
Ceniceros and Fredrickson\,\cite{ceniceros2004numerical} devised a class of efficient
semi-implicit schemes for solving the SCFT equations using the
asymptotic expansion technology. Later, Jiang et al.\,\cite{jiang2015analytic} have
extended these algorithms to the SCFT calculations for multicomponent polymer systems.

A generic strategy of theoretical studying phase behavior of
complex block copolymer systems includes two steps\,\cite{li2010real, xu2013strategy}.
The first step involves an efficient strategy to produce a
library of possible candidate structures.
In the second step, the
candidate structures are used as initial conditions in the more
accurate methods to compute free energies, which are used to
construct phase diagrams of the systems.
In this work, we apply this strategy to examine the phase
behavior of ABC star triblock copolymers using the SCFT.
Specifically, the strategies developed in our previous work\,\cite{xu2013strategy} are used as a screening technique to obtain
candidate structures as many as possible.
These strategies include (1) knowledge from previous experiments and theories; (2)
knowledge from related systems, for example, diblock
copolymers; (3) combination and interpolation of known
structures; and (4) random initial configurations.
Using these candidate structures as initial conditions,
a fourth-order pseudospectral method combined with Anderson
mixing method is employed to study the stability of ordered phases.
To model the asymmetric interacting experimental systems of ISP and ISF, the interaction parameters of
$\chi_{AC}N>\chi_{BC}N=\chi_{AB}N$ are used in our study.
It should be emphasized that, to broaden the scope of the
research, the 2D and 3D ordered phases are included in our calculations.
The resulting free energies of the different ordered phases are
used to construct phase diagrams.

\section{Theory and Methods}
\label{sec:theory}

We consider an incompressible melt of $n$ ABC star triblock copolymers
with a degree of polymerization $N$ in a finite volume
of $V$. The chain lengths, or compositions, of A, B, and C blocks are $f_A N$,
$f_B N$, and $f_C N$ ($f_A + f_B + f_C = 1$), respectively.
A characteristic length of the copolymer chain can be defined by
the radius of gyration, which is used as the unit of length, so that
all spatial lengths are presented in units of $R_g$.
Within the mean-field approximation to statistical mechanics of
the Edwards model of polymers, at a temperature T, the free
energy functional $F$ per chain of the Gaussian triblock
copolymer melt is
\begin{align}
  \frac{F}{nk_B T} = -\log Q + \frac{1}{V}\int
  d\mathbf{r}\,\Big\{ \sum_{\alpha\neq\beta}\chi_{\alpha\beta}N \phi_\alpha(\mathbf{r}) \phi_\beta(\mathbf{r})
   -\sum_{\alpha} w_\alpha(\mathbf{r}) \phi_\alpha(\mathbf{r}) -
   \eta(\mathbf{r})[1-\sum_{\alpha} \phi_\alpha(\mathbf{r})]\Big\},
  \label{eqnscftenergy}
\end{align}
where $\alpha, \beta \in \{A,B,C\} $ are the block labels,
$\phi_\alpha$ is the monomer density of the $\alpha$-blocks,
and $Q$ is the partition function of one star block copolymer
chain in the mean field, $w_\alpha(\mathbf{r})$, produced by the surrounding chains.
$\eta(\mathbf{r})$ is the Lagrange field to ensure the local incompressibility.
The interactions between the three chemically distinct monomers
are characterized by three Flory-Huggins interaction parameters
multiplied by polymerization degree, i.e.,
$\chi_{BC}N$, $\chi_{AC}N$, and $\chi_{AB}N$.
First-order variations of the free-energy functional with respect
to the monomer densities and the conjugate fields subjected to the
incompressible condition lead to the following set of SCFT equations
\begin{align}
	w_A(\mathbf{r})& =
	\chi_{AB}N\phi_B(\mathbf{r}) +
	\chi_{AC}N\phi_C(\mathbf{r}) + \eta(\mathbf{r}),
	\label{eqscftwa}
    \\
	w_B(\mathbf{r}) &=
	\chi_{AB}N\phi_A(\mathbf{r}) +
	\chi_{BC}N\phi_C(\mathbf{r}) + \eta(\mathbf{r}),
	\label{eqscftwb}
    \\
	w_C(\mathbf{r}) &=
	\chi_{AC}N\phi_A(\mathbf{r}) +
	\chi_{BC}N\phi_B(\mathbf{r}) + \eta(\mathbf{r}),
	\label{eqscftwc}
	\\
	1 &= \phi_A(\mathbf{r})+\phi_B(\mathbf{r})+\phi_C(\mathbf{r}),
	\label{eqscftincompressibility}
	\\
	\phi_A(\mathbf{r}) &= \frac{1}{Q} \int_0^{f_A}
	ds\, q_A(\mathbf{r}, s) q_A^\dag(\mathbf{r}, f_A-s),
	\label{eqscftphiA}
	\\
	\phi_B(\mathbf{r}) &= \frac{1}{Q} \int_0^{f_B}
	ds\, q_B(\mathbf{r}, s) q_B^\dag(\mathbf{r}, f_B-s),
	\label{eqscftphiB}
	\\
	\phi_C(\mathbf{r}) &= \frac{1}{Q} \int_0^{f_C}
	ds\, q_C(\mathbf{r}, s) q_C^\dag(\mathbf{r}, f_C-s),
	\label{eqscftphiC}
	\\
	Q &= \frac{1}{V} \int d\mathbf{r}\, q_K(\mathbf{r}, s)
	q_K^\dag(\mathbf{r}, f_K-s), ~~~\forall s, K.
	\label{eqscftQ}
\end{align}
In these expressions, the functions $q_K(\mathbf{r}, s)$
and $q_K^\dag(\mathbf{r}, s)$ ($K\in\{A,B,C\}$)
are the end-integrated segment distribution functions, or
propagators, representing the probability of finding the $s$-th
segment at a spatial position $\mathbf{r}$. These propagators
satisfy the modified diffusion equations (MDEs)
\begin{equation}
	\begin{aligned}
		&\frac{\partial}{\partial s}q_K(\mathbf{r}, s) = \nabla^2_{\mathbf{r}}
		q_K(\mathbf{r}, s) - w_K q_K(\mathbf{r}, s),
		~~~ q_K(\mathbf{r}, 0) = 1, ~~s\in[0,f_K],
		\\
		&\frac{\partial}{\partial s}q_K^\dag(\mathbf{r}, s) = \nabla^2_{\mathbf{r}}
		q_K^\dag(\mathbf{r}, s) - w_K q_K^\dag(\mathbf{r}, s),
		~~~ q_K^\dag(\mathbf{r}, 0) = q_L(\mathbf{r},
		f_L) q_M(\mathbf{r}, f_M),
	\end{aligned}
	\label{eqscftqplus}
\end{equation}
where $(K L M)\in\{(ABC),(BCA),(CAB)\}$.
Numerically solving
these SCFT equations involves an iterative procedure starting
with an initial configuration of the fields $w_K(\mathbf{r})$ with $K\in{A,B,C}$.
The MDEs (Eqn.\,\eqref{eqscftqplus}) are then solved to
obtain the propagators, which are used to compute the
densities $\phi_K(\mathbf{r})$ and to update the mean fields $w_K(\mathbf{r})$.
The iteration is continued until these mean fields and densities
are self-consistent such that they satisfy the SCFT equations
within a prescribed numerical accuracy.

There are two main steps in studying the phase behavior of
block copolymers. The first step is to obtain possible candidate
structures as many as possible. Several strategies of exploring ordered
phases have been proposed in our previous works\,\cite{jiang2010spectral, xu2013strategy}.
Beyond the random initial values, these approaches include (1) knowledge from experiments and
theories, such as small-angle X-ray scattering images, space group theory for periodic structures; (2)
knowledge from related systems; (3) combination and interpolation of known structures.
Using these diverse strategies of initialization, a large number
of ordered phases can be generated as the solutions of the SCFT equations.
In addition, the possible candidate structures from previous theoretical and experimental studies are considered in our calculations.
The second step is to identify the stability of these phases
by comparing their free energies using efficient numerical methods.
In this work, we combine an improved pseudosepectral method with
Anderson mixing algorithm to solve the SCFT equations for periodic
block-copolymer morphologies. A fourth-order accurate
Adams-Bashford scheme\,\cite{cochran2006stability} is used to
discretize the MDEs. The initial values required to apply this
formula are obtained using a special extrapolation
method\,\cite{ranjan2008linear}, based on
the second-order operator-splitting scheme\,\cite{rasmussen2002improved}.
A modified integral formula for closed interval is chosen to solve the
integrated equations \eqref{eqscftphiA}-\eqref{eqscftphiC}
that can guarantee fourth-order precision in $s$-direction
whether the number of discretization points is even or
odd\,\cite{press1992numerical}.
To ensure the accuracy, we require that these substeps of contour
length $s$ are smaller than $0.01$ in the fourth-order accurate scheme.
The FFT is used to translate the data between real- and
Fourier-space in the pseudospectral method.

To obtain the equilibrium morphologies of SCFT equations,
iterative methods shall be required to update the conjugate
fields. For this step we choose the Anderson mixing algorithm,
firstly proposed by Anderson\,\cite{anderson1965iterative}, then
introduced into polymer theoretical calculations by Schmid and
M{\"u}ller\,\cite{schmid1995quantitative}, Thompson et al.\,\cite{thompson2004improved}
Owing to the local convergence of this method, we use the simple mixing method alone at
the start of the algorithm to obtain better initial values for
the fields, followed by Anderson mixing approach alone to accelerate the
convergent procedure to the prescribed accuracy in the fields.
The Anderson mixing method requires the fields of previous $n$ steps when update new fields.
Using the previous $n$ step fields, the Anderson mixing method
produces an $n$-order linear equations system from a least square problem.
Then the new fields will be obtained by a combination of the
fields of previous $n$ steps, with the solution of linear system being the weight factors.
For relatively simpler systems, such as diblock copolymers,
the Anderson mixing algorithm can significantly reduce the
required number of iterations with few histories\,\cite{thompson2004improved, matsen2009fast}.
For more complex situation, a larger $n$ shall be taken to
update fields to accelerate the convergent procedure.
We find that assembling the $n$-order linear system
spends more computation time than solving this linear system.
When $n$ becomes too large, it will slow the iteration.
Here we overcome this problem by rearranging the elements of the
$n\times n$ matrix. The technical details can be found in the Appendix section.
By using our approach, the Anderson mixing method can
robustly converge without slowing the SCFT iterations.
In practice, we use the available histories of $50$ steps.

Here we only consider the periodic structures, therefore,
periodic boundary conditions are imposed on each direction.
In our calculations, all spatial functions are all expanded in
terms of plane waves.
For 2D morphologies, a square box is simulated. $128\times128$
plane-wave basis functions are used to discretize the 2D box.
For 3D structures, we use a cubic unit cell in most of our
calculations. The number of plane-wave basis functions is $32\times32\times32$.
The size of computation box plays an important role in
determining the stability of ordered phases.
For a given phase, its free energy is minimized with respect to
the box sizes by the steepest descent approach
coupled with solving the SCFT equations\,\cite{jiang2010spectral}.

Based on the discussion above, we summarize the iteration
procedure by sketching the numerical recipes:
\begin{description}
 \item [Step 1] Starting from the given initial conditions and
	 computational box, the fourth-order pseudosepectral method combined with
	 Anderson mixing method is applied to obtain the ordered phases
	 when the field's change is smaller than the prescribed error $\varepsilon_1$.
 \item [Step 2] Optimizing the size of unit cell by minimizing the
	 free energy with the steepest descent method.
 \item [Step 3] Goto \textbf{Step 1} until the free energy change is smaller
	 than a given error $\varepsilon_2$.
\end{description}
To ensure enough accuracy, in our implementation, each
calculation is terminated until the field's change
(defined in Appendix) at each iteration is reduced to
$\varepsilon_1=10^{-6}$ (corresponding to a free energy change of
about $10^{-7}$), and $\varepsilon_2=10^{-6}$. For the cases of
$\chi_{AC}N=50.0$, $\chi_{AB}N=\chi_{BC}N=30.0$ in the current work,
the relative values of free energy among different phases in determining the
phase boundary are from $O(10^{-2})$ to $O(10^{-4})$.
Therefore our numerical resolution is adequately accurate for
constructing phase diagrams.

\section{Results and Discussion}
\label{sec:rslt}

The previously theoretical studies mainly focus on the equally interacting ABC
star triblock copolymer systems, i.e.,
$\chi_{BC}N=\chi_{AC}N=\chi_{AB}N$\,\cite{gemma2002monte, huang2008morphological, li2010real, zhang2010scft, xu2013strategy}.
However, under experimental circumstances as mentioned above, many ABC star triblock
copolymers with asymmetric interactions have been synthesized to observe their phase behaviors, such
as ISP\,\cite{matsushita2010jewelry} and ISF\,\cite{nunns2013synthesis} star triblock copolymers.
In particular, the interaction parameters of
the above systems satisfy the relationship
$\chi_{AB}N\approx \chi_{BC}N<\chi_{AC}N$.
In order to make a meaningful comparison of the theoretical study and
experimental investigation, the interaction parameters are chosen in the
calculations so that they are appropriate for these experimental systems.
Although accurate values of the Flory-Huggins parameters are not
available in the literatures, qualitative interaction strengths are known to follow the order
$\chi_{IS} \approx\chi_{SP} < \chi_{IP}$\,\cite{takano2005mesoscopic, matsushita2010jewelry},
$\chi_{SF} \approx\chi_{IS} < \chi_{IF}$\,\cite{nunns2013synthesis}.
In what follows we choose asymmetric interaction parameters,
$\chi_{AC}N=50.0$, $\chi_{AB}N=\chi_{BC}N=30.0$, and equal statistical segment lengths.
This is a rough approximation to ISP and ISF copolymers, with the
important difference that, in the idealized system, we neglect
the small differences between $\chi_{AB}N$ and $\chi_{BC}N$.

As mentioned above, there are two main steps in studying the phase behavior of block
copolymers. The first step is to obtain as many possible candidate structures
as possible. The second step is to identify the stability of these patterns by comparing
their free energies and construct the phase diagrams.
In order to further analyse the stability of candidate patterns,
it is helpful to split the free energy into two parts: internal
($U$) and entropic ($-TS$). The internal and entropic
contributions to the free energy can be expressed as\,\cite{matsen2002standard}
\begin{equation}
\begin{aligned}
	\frac{U}{nk_B T}&=\frac{1}{V}\int d\mathbf{r}\,
	\Big\{
	\sum_{\alpha\neq\beta}\chi_{\alpha\beta}N \phi_\alpha(\mathbf{r}) \phi_\beta(\mathbf{r})
	\Big\},
	\\
	-\frac{S}{nk_B T}&=-\log Q-\frac{1}{V}\int d\mathbf{r}\,
	\Big\{
   \sum_{\alpha} w_\alpha(\mathbf{r}) \phi_\alpha(\mathbf{r})
	\Big\},
\end{aligned}
	\label{eqnenergysplit}
\end{equation}
where $\alpha, \beta \in \{A,B,C\} $ are the block labels.

\subsection{Candidate Patterns}

\begin{figure*}[!hbpt]
  \centering
\includegraphics[width=0.7\columnwidth]{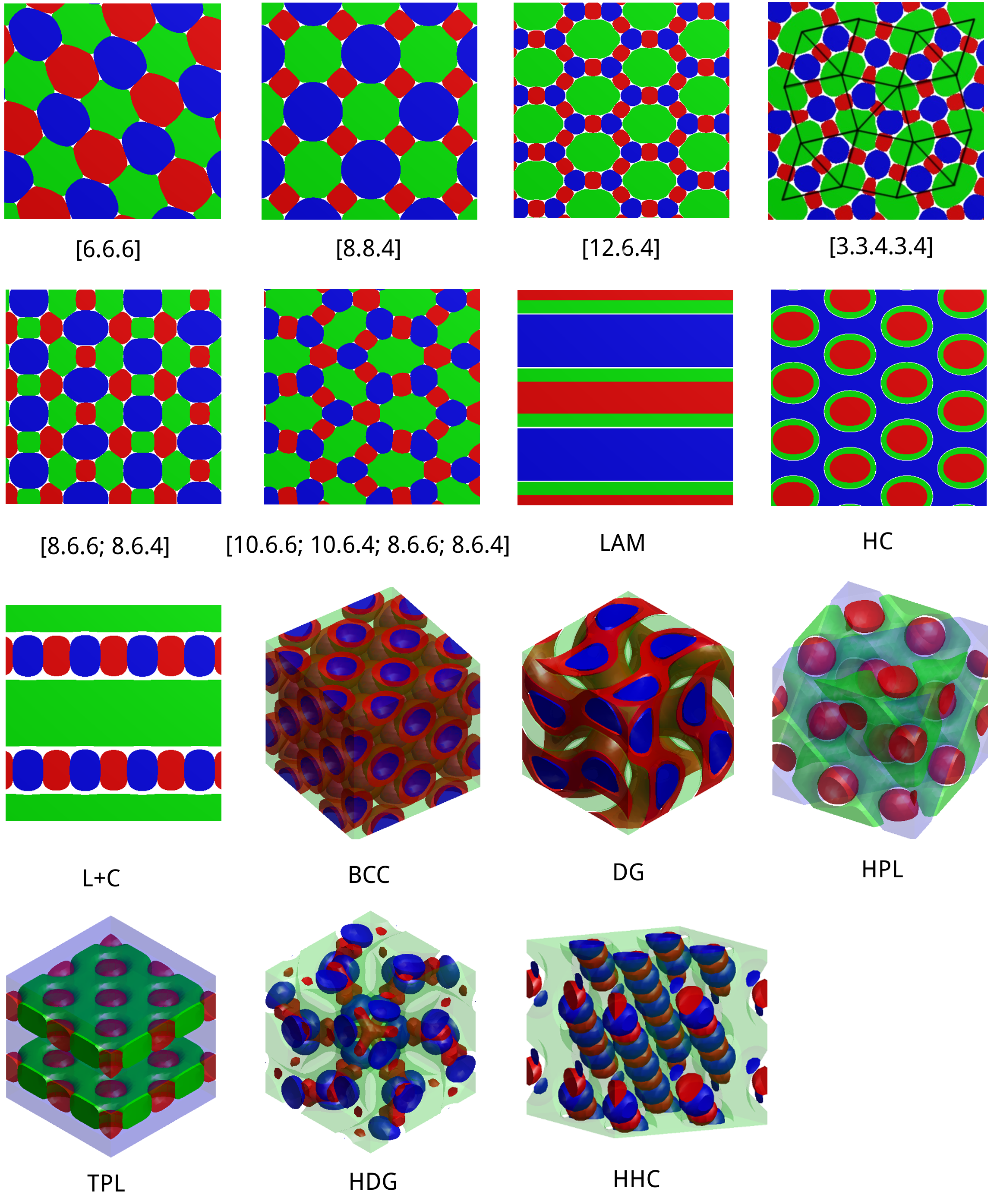}
  \caption{Ordered phases of ABC star triblock copolymers
  obtained using the SCFT calculations with $\chi_{AC}N=50.0$,
  $\chi_{AB}N=\chi_{BC}N=30.0$. The colors of red,
  green, and blue, indicate the regions where the most components
  are A, B, and C, respectively.}
  \label{fig:phases}
\end{figure*}
Using different initialization procedures,
a large number of candidate structures have been obtained in our
previous work\,\cite{xu2013strategy}.
Here we only present the stable phases in the case of
$\chi_{AC}N=50.0$, $\chi_{AB}N=\chi_{BC}N=30.0$, as shown in Fig.\,\ref{fig:phases}.
Among these candidate phases,
several polygon tiling patterns, or the cylindrical structures, with
translation invariant along the third direction are found as 2D phases.
These polygon patterns include [6.6.6],
[8.8.4], [12.6.4], [3.3.4.3.4], [8.6.6;8.6.4],
[10.6.6;10.6.4;8.6.6;8.6.4].
The first pattern is designated
as [6.6.6] because each vertex in this tiling is surrounded by
three hexagonal polygons.
Similarly the second and third patterns are named
[8.8.4] and [12.6.4], respectively.
The fourth pattern looks more complex. There are two types of
vertices: one is surrounded by 10-gon, 8-gon and 4-gon, whereas
the other is formed by a decagon, a hexagon and a tetragon. From
our naming convention, it should be named
[10.8.4;10.6.4]. However, with the help of
triangles and squares of hypothetical tiling, as
Fig.\,\ref{fig:phases} shows
the superimposed tiling on a schematic drawing, it is noted
all the meeting vertices are surrounded by three regular triangles
and two squares, which is one of the Archimedean tilings. In
order to compare with experimental results, we encode it as
[3.3.4.3.4]\,\cite{takano2005mesoscopic}. The fifth pattern also possesses two kinds
of vertices: one is surrounded by 8-gon, 6-gon, and 4-gon,
whereas the other
is formed by an 8-gon and two 6-gons. Consequently, this pattern
is designated as [8.6.4;8.6.6].  Similarly, the fifth pattern has
four kinds of vertices, therefore, it is named [10.6.6;10.6.4;8.6.6;8.6.4].
Besides these polygonal phases, additional three 2D structures, i.e., three color lamellae
(LAM), the core-shell cylinders (HC), and hierarchical
cylinders-in-lamella phases (L+C) are also obtained in our simulations.
At the same time, a series of 3D structures are obtained.
From the morphology of patterns, these 3D patterns can be classified into
core-shell phases and hierarchical structures.
The former includes core-shell spheres in body-centered-cubic
lattice (BCC) and  core-shell double-gyroid phases (DG).
The hierarchical patterns consist of two kinds of hierarchical cylinders
packed hexagonally (HHC),
two kinds of cylinders-in-lamella phases, with the cylinders being packed hexagonally (HPL) and
tetragonally (TPL), and hierarchical gyroid phases (HDG).
Note that due to the equal interaction parameters of $\chi_{AB}N=\chi_{BC}N$, these
structures have their mirror phases along the phase path of isopleth $f_A=f_C$.

\subsection{Phase Behavior of $\mathrm{A_{1.0}B_{1.0}C_{x}}$ Star Triblock Copolymers}

\begin{figure}[!htbp]
  \centering
\includegraphics[width=.48\columnwidth]{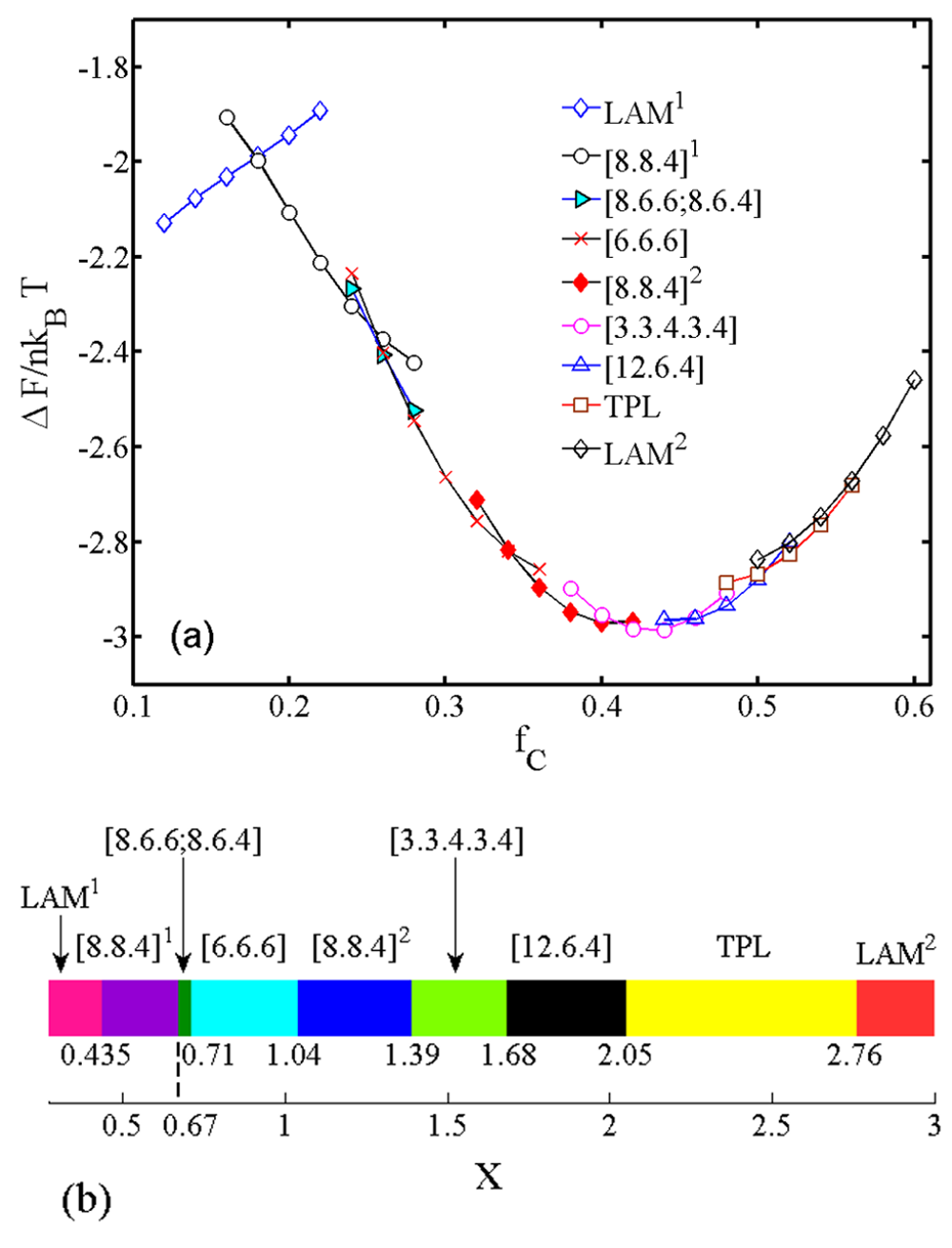}
	  \caption{(a) Free energy differences from the value of the
	  homogeneous phase as a function of the volume fraction of
	  C composition for ABC star triblock copolymers with symmetric A and
	  B arms. (b) Phase stability regions as a function of the
	  arm-length ratio of $x=f_C/f_A$ ($f_A=f_B$).
	  Note that in the $[8.8.4]^1$ phase, the minority C blocks
	  form the 4-coordinated domains, and blocks A and B
	  alternatively form 8-coordinated microdomains.
	  While in the $[8.8.4]^2$ morphology, the A and C blocks
	  form the 8-coordinated polygons, and B blocks form the
	  domains with 4-coordinations.
	  In one periodicity, the $\mathrm{LAM^1}$ phase has the CACB lamellar sequence, whereas
	  the $\mathrm{LAM^2}$ structure has the BABC layers.
	  }
\label{fig:11x}
\end{figure}

In a series of experiments\,\cite{takano2005mesoscopic,
hayashid2006archimedean, takano2007composition, nunns2013synthesis},
two of the three arms are kept to be of equal
length and the arm-length ratio is expressed as $1.0:1.0:x$.
Motivated by these experiments, we start with the calculation of the phase stability along this
phase path. Here we assume that A and B arms have equal
length and the C arm holds the arm-length ratio of $x = f_C/f_A$.
The free energy differences of candidate phases
from the value of the homogeneous phase as a function of
the volume fraction of $f_C$ are given in Fig.\,\ref{fig:11x}\,(a).
The phase stability regions as a function of $x$ are presented in Fig.\,\ref{fig:11x}\,(b).
In Fig.\,\ref{fig:11x}\,(a), the free energies of some metastable
phases along this path are not shown, such as that of [10.6.6;10.6.4;8.6.6;8.6.4]
in the region of $0.32\leq f_C \leq 0.46$, where it has higher
free energy than [6.6.6], or $[8.8.4]^2$, or [3.3.4.3.4].
With the increase of $x$, the phase sequence is $\mathrm{LAM^1}$
$\rightarrow$ $[8.8.4]^1$ $\rightarrow$  [8.6.6;8.6.4]
$\rightarrow$ [6.6.6] $\rightarrow$ $[8.8.4]^2$ $\rightarrow$ [3.3.4.3.4] $\rightarrow$
[12.6.4] $\rightarrow$ TPL $\rightarrow$ $\mathrm{LAM^2}$.
The corresponding stable regions are $x\leq0.435$ ($\mathrm{LAM^1}$),
$0.435\leq x\leq 0.67$ ($[8.8.4]^1$), $0.67\leq x\leq 0.71$ ([8.6.6;8.6.4]), $0.71\leq
x\leq 1.04$ ([6.6.6]), $1.04\leq x\leq 1.39$ ($[8.8.4]^2$),
$1.39\leq x\leq 1.68$ (3.3.4.3.4), $1.68\leq x\leq 2.05$ ([12.6.4]),
$2.05\leq x\leq 2.76$ (TPL), and $x\geq2.76$ ($\mathrm{LAM^2}$), respectively.
When $f_C$ is small, in one period, the lamellar structure
of $\mathrm{LAM}^1$ of CACB type includes two thick A and B layers, and two thin C layers.
The symbols of $[8.8.4]^1$ and $[8.8.4]^2$ are used to distinguish
the 4-coordinated polygons segregated by different block.
In the $[8.8.4]^1$ phase, the minority C-arms
form the 4-coordinated domains, and blocks A and B alternatively form 8-coordinated polygons.
While the unit cell of $[8.8.4]^2$ contains one 4-coordinated
B-domain, 8-coordinated and 4-coordinated microdomains formed by blocks A and C, respectively.
From the phase path, in tiling patterns, the
coordination number of C-domains is proportional to the volume
fraction of $f_C$. With the increment of $f_C$ from $0.18$ to $0.50$,
the coordinations of C-domains change from $4$, to $6$, to $8$,
then to $10$ (in the [3.3.4.3.4] phase), finally to $12$.
Further increasing $f_C$, the asymmetry of A (B) arm and C
arm results in arrangement of junctions on a curve line.
Then a 3D structure of TPL appears when $0.51 \leq f_C \leq 0.58$.
\begin{figure}[!htbp]
  \centering
\includegraphics[width=.48\columnwidth]{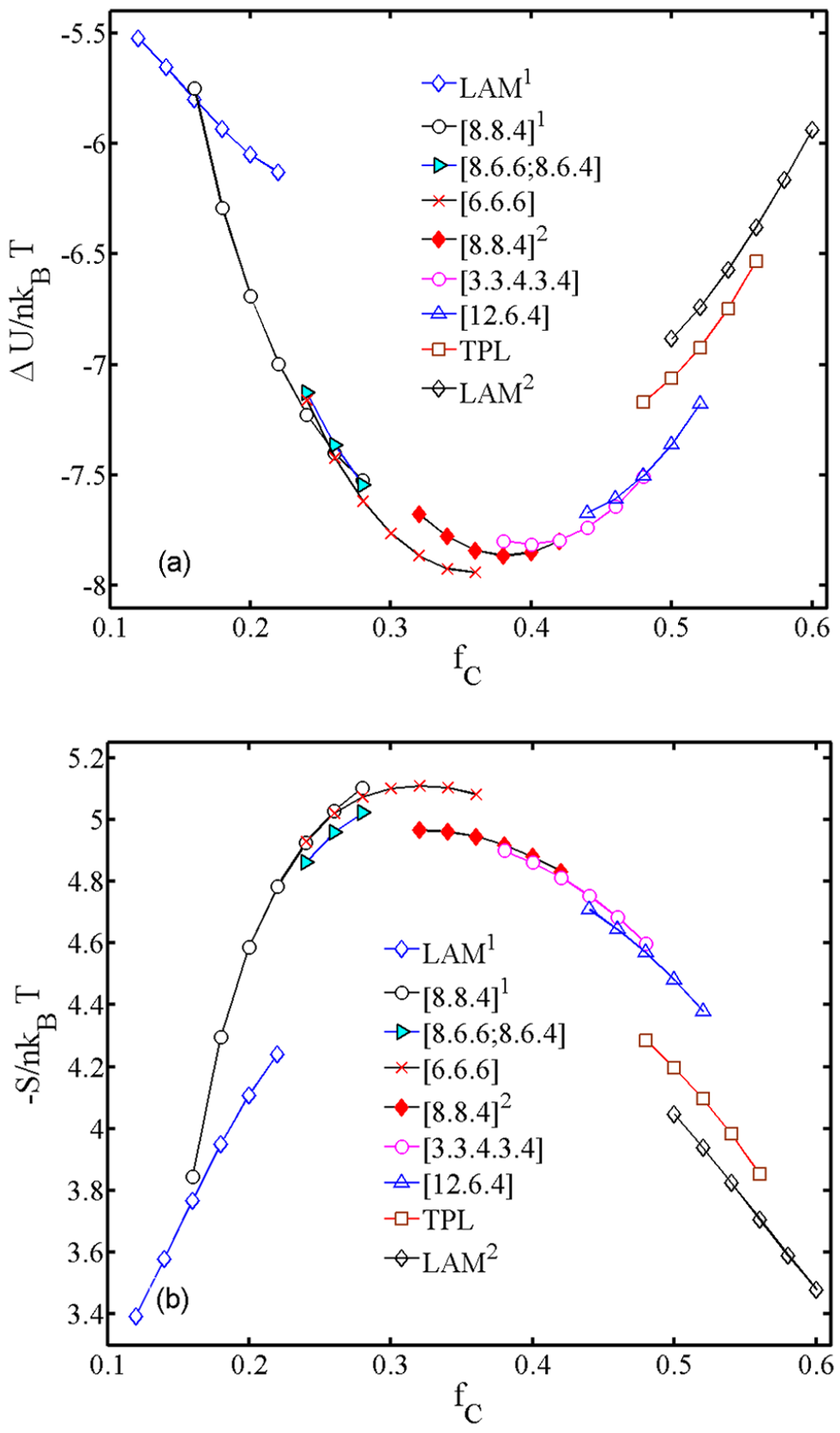}
	  \caption{(a) Internal energy of $\Delta U/nk_B T$ and
	  (b) entropic energy of $-S/nk_B T$ of various structures as a
  function of $f_C$ on the phase path of $f_A=f_B$.
  The morphologies of $[8.8.4]^1$, $[8.8.4]^2$, $\mathrm{LAM^1}$
  and $\mathrm{LAM^2}$ are explained in Fig.\,\ref{fig:11x}.
  }
\label{fig:energysplit11x}
\end{figure}
When $f_C>0.58$, the lamellar phase $\mathrm{LAM}^2$ is
stable in which the arrangement manner is BABC type in one period,
including one thick C layer and three thin BAB layers.

We can analyze the stability of these ordered phases
after splitting the SCFT energy functional into the internal and
entropic parts (see Eq.\,\eqref{eqnenergysplit}).
Fig.\,\ref{fig:energysplit11x} gives the internal energy
$\Delta U/nk_B T$, subtracted by that of the homogeneous
phase\,\cite{li2010real}, together with the entropic energy of
$-S/nk_B$, as a function of $f_C$.
From Fig.\,\ref{fig:energysplit11x}\,(a), we can
find that $\mathrm{LAM^1}$ (CACB layers) has very high internal energy at small $f_C$ which is induced by
the penetrations of A and B arms through C domains\,\cite{li2010real}. At the
same time, the lamellae of $\mathrm{LAM^1}$ is favorable from the aspect of entropic energy,
because the A and B blocks can get the largest entropy in the lamellar structure
when they have equal large lengths compared with the C blocks.
The combination of the two contributions makes the $\mathrm{LAM^1}$ be
the stable phase when $f_C \leq 0.18 $. When increasing $f_C$, the
internal energy becomes dominant and
the stable phase transfers from $\mathrm{LAM^1}$ to the
cylindrical structures where the arm penetrations in dissimilar phase regions are diminished.
When the stable phase transfers from $[8.8.4]^2$ to [3.3.4.3.4]
(also termed as [10.8.4; 10.6.4]),
the entropic energy plays a dominant role on the stability.
It is attributed to the increment of C arm which tends to form large C-domains.
The enlarged C-domains have an opportunity to meet much more A-domains
which will increase the internal energy due to the
largest interaction parameter $\chi_{AC}N$.
On the other hand, as the arm C increases, the A and B arms become shorter.
The A and B arms can freely stretch in their
rich domains which can greatly reduce the entropic energy.
The explanation is also available to the appearance of
[12.6.4], TPL as stable phases.
Further increasing $f_C$ to $0.58$, the lamellar phase of $\mathrm{LAM^2}$ (BABC layers)
is stable again. The reason is similar to that of the stability
of $\mathrm{LAM^1}$ phase when $f_C$ is smaller than $0.18$.

Along this phase path, Matsushita and co-workers\,\cite{hayashid2006archimedean}
synthesized a set of $\mathrm{I_{1.0}S_{1.0}P_{x}}$ copolymers and observed a number of ordered structures.
Four ISP star triblock copolymers with volume ratios
of 1.0:1.0:0.7, 1.0:1.0:1.2, 1.0:1.0:1.3, and 1.0:1.0:1.9 were investigated.
The cylindrical structures of
$\mathrm{I_{1.0}S_{1.0}P_{0.7}}$,
$\mathrm{I_{1.0}S_{1.0}P_{1.2}}$,
$\mathrm{I_{1.0}S_{1.0}P_{1.3}}$, and $\mathrm{I_{1.0}S_{1.0}P_{1.9}}$
are $[6.6.6]$, $[8.8.4]$, $[3.3.4.3.4]$, and $[12.6.4]$, respectively.
From our simulations,
the resulting phase behavior is in good agreement with these
experimental measurements when $\mathrm{0.7<x<2.0}$.
At higher asymmetries, Takano et al.\,\cite{takano2007composition}
observed an L+S phase in $\mathrm{I_{1.0}S_{1.0}P_{0.2}}$ system, and
an L+C phase in $\mathrm{I_{1.0}S_{1.0}P_{3.0}}$ and
$\mathrm{I_{1.0}S_{1.0}P_{4.9}}$ copolymers.
However, in our calculations, the LAM phase dominates the regions.
This might be attributed to the relatively weak interaction parameters of
$\chi_{AB}N$ or $\chi_{BC}N$ in our simulations so that a
further segregation between components A and B (or B and C) can not occur.
In 2007, Hayashida et al.\,\cite{hayashida2007hierarchical} have
discovered a cylinders-in-lamella phase in which the stacking
manner of the cylinders seems to be random in the experiments of
ISP star triblock copolymer melts.
In our calculations, the cylinders-in-lamella, TPL, is found to be
stable when $0.51 \leq f_C \leq 0.56$, in which the cylinders
are stacking tetragonally.
The discrepancy is attributed to the thermodynamic fluctuations which
may affect the arrangement of cylinders under experimental
circumstances. While within the mean-field level theory, the
fluctuations have been neglected.

Along the similar phase path, Nunns et al.\,\cite{nunns2013synthesis}
synthesized a set of ISF star triblock copolymers and observed
the polygonal tilings [8.8.4] and [12.6.4].
Our resulting phase behavior is generally consistent with the ISF experiments.
A deviation between our theoretical results and experiments is
that the [12.6.4] was observed in $\mathrm{I_{0.40}S_{0.37}F_{0.23}}$ system.
The discrepancy can be attributed to the different interactions,
and different monomer sizes.

Our computational results also agree with the previous
theoretical calculations\,\cite{gemma2002monte, zhang2010scft, li2010real}.
Among these works, Zhang et al.\,\cite{zhang2010scft} considered 2D
tiling patterns and chose the asymmetric Flory-Huggins interaction
parameters of $\chi_{AC}N=37.0$, $\chi_{AB}N=\chi_{BC}N=25.0$ to
model the system of ABC star triblock copolymers.
A 1D phase diagram of the star triblock
copolymers $\mathrm{A_{1.0}B_{1.0}C_{x}}$ was obtained. With the
increase of $x$, the phase transition changes from $[8.8.4]^1$ to
$[8.8.4]^2$, then to [8.6.6;8.6.4] and finally to [12.6.4].
The corresponding stable regions are $0.50\leq x\leq 0.86$,
$0.94\leq x\leq 1.33$, $x\approx 1.45$ and $1.57\leq x\leq 2.00$, respectively.
The phase behaviors of $[8.8.4]^1$, $[8.8.4]^2$, and
[12.6.4] are qualitatively consistent with our results.
There are some discrepancies between their results and our computer simulations.
In their phase diagram, the Archimedean tiling of
[3.3.4.3.4] is not included in their simulations,
the order pattern [6.6.6] is metastable,
and the stable area of [8.6.6;8.6.4] is different from ours.
The discrepancies can be attributed to two aspects. The first one is that more candidate
structures are involved in our simulations. The second one is
the difference of the Flory-Huggins interaction parameters.

\subsection{Phase Behavior of $\mathrm{A_{1.0}B_{x}C_{1.0}}$ Star Triblock Copolymers}

On the above phase path, many candidate structures, such as 2D
polygonal phases [10.6.6;10.6.4;

\begin{figure}[!htbp]
  \centering
\includegraphics[width=.48\columnwidth]{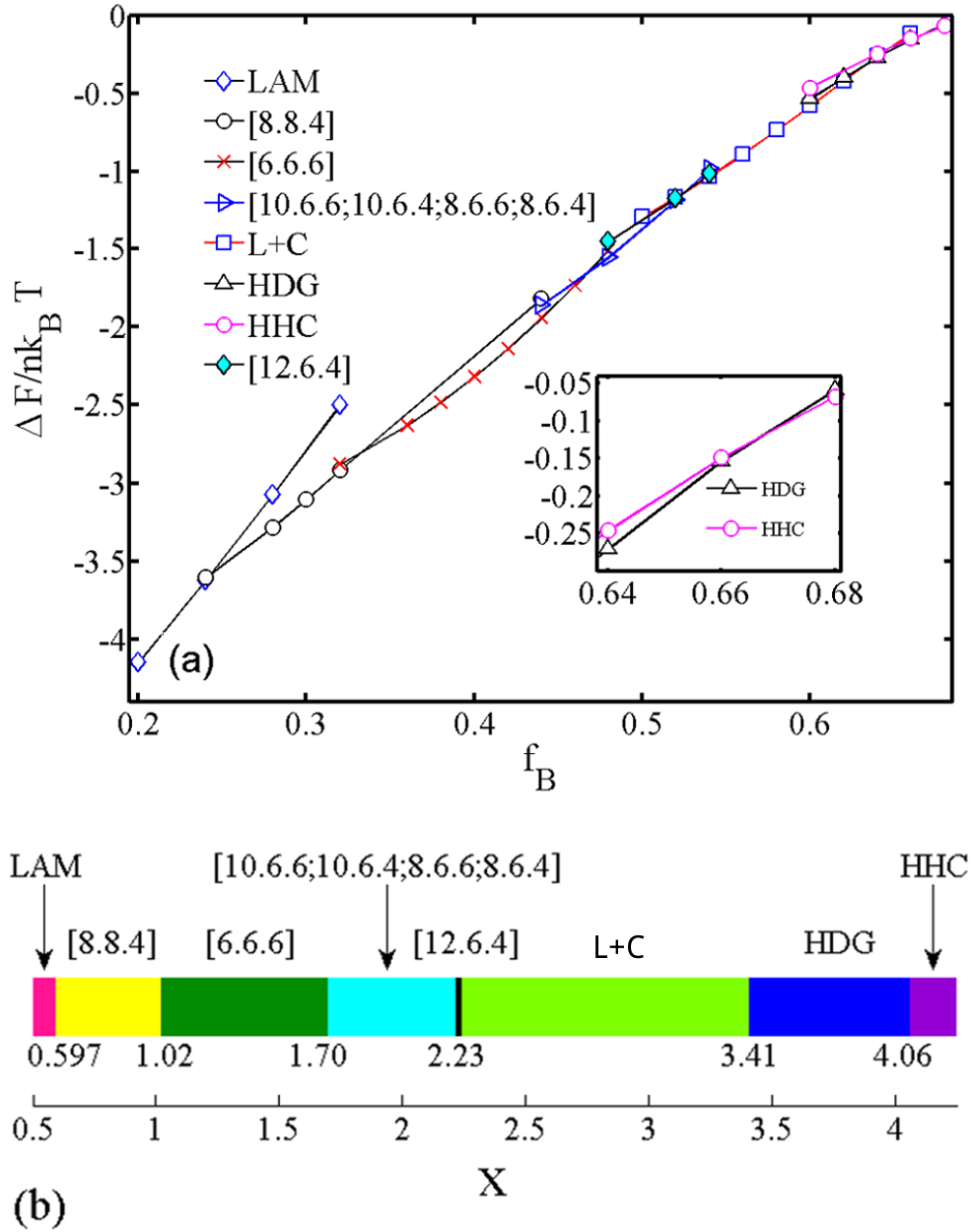}
	  \caption{(a) The free energy differences of various candidate phases,
	  from the value of the homogeneous phase, as a function of
	  $f_B$ for ABC star triblock copolymers with symmetric A and C arms.
	  (b) Phase stability regions as a function of the arm-length ratio of
	  $x=f_B/f_A$ ($f_A=f_C$) of ABC star triblock
	  copolymers with asymmetric interactions, $\chi_{AC}N=50.0$,
	  $\chi_{AB}N=\chi_{BC}N=30.0$ as a function of the
	  arm-length ratio of $x=f_B/f_A$.}
\label{fig:1x1}
\end{figure}

\noindent 8.6.6;8.6.4] and 3D hierarchical
structures of HHC, HDG, do not appear.
To obtain the stability regions of the 2D and 3D
phases which have been observed in experiments, we turn to
another phase path of isopleth $f_A=f_C$, i.e.,
by fixed equal length of arms A and C and the arm-length ratio of $1.0:x:1.0$
with an increment of volume fraction $f_B$ of $0.01$.
The free energy difference from the
value of the homogeneous phase as a function of $f_B$ varying
from $0.20$ to $0.68$ is plotted in Fig.\,\ref{fig:1x1}\,(a).
The phase stability regions as a function of $x$ are presented in
Fig.\,\ref{fig:1x1}\,(b).
Along this phase path, the phase sequence with increasing
$f_B$ is LAM $\rightarrow$ [8.8.4] $\rightarrow$ [6.6.6]
$\rightarrow$ [10.6.6;10.6.4;8.6.6;8.6.4] $\rightarrow$ [12.6.4]
$\rightarrow$ L+C $\rightarrow$ HDG $\rightarrow$ HHC.
The stable region of LAM phase is $f_B\leq 0.24$.
At the center part of the phase path, the chain lengths of three
arms are close to one another, junction points are aligned
on a straight line, and hence 2D tiling patterns can be formed.
The stability regions of 2D cylindrical phases are
$0.24\leq f_B \leq 0.33$ ([8.8.4]), $0.33\leq f_B \leq 0.46$ ([6.6.6]), $0.46 \leq f_B \leq
0.52$ ([10.6.6;10.6.4;8.6.6;8.6.4]), $f_B\approx 0.53$
([12.6.4]), respectively.
From the phase path, we can also find that, in polygonal patterns, the
coordinations of B-domains are proportional to the volume
fraction of $f_B$. With the increase of $f_B$ from $0.24$ to $0.53$,
the coordination number of B-domains goes through a
gradual change from $4$ ($\mathrm{[8.8.4]^1}$), $6$ ([6.6.6]),
$8$ or $10$ ([10.6.6;10.6.4;8.6.6;8.6.4] tiling), to $12$
([12.6.4]).

Further increasing $f_B$, the A and C arms become shorter.
Also, owing to the largest interaction parameter $\chi_{AC}N$,
a further segregation between two minority blocks A and C occurs within the
large-length-scale phase, and the system can form some hierarchical morphologies.
The interesting hierarchical patterns include 2D L+C phase, and 3D patterns of HDG, HHC, and their stability regions
are $0.54\leq f_B \leq 0.63$ (L+C), $0.63\leq f_B \leq 0.67$ (HDG), $f_B \geq 0.67$ (HHC), respectively.

\begin{figure}[!htbp]
  \centering
\includegraphics[width=.48\columnwidth]{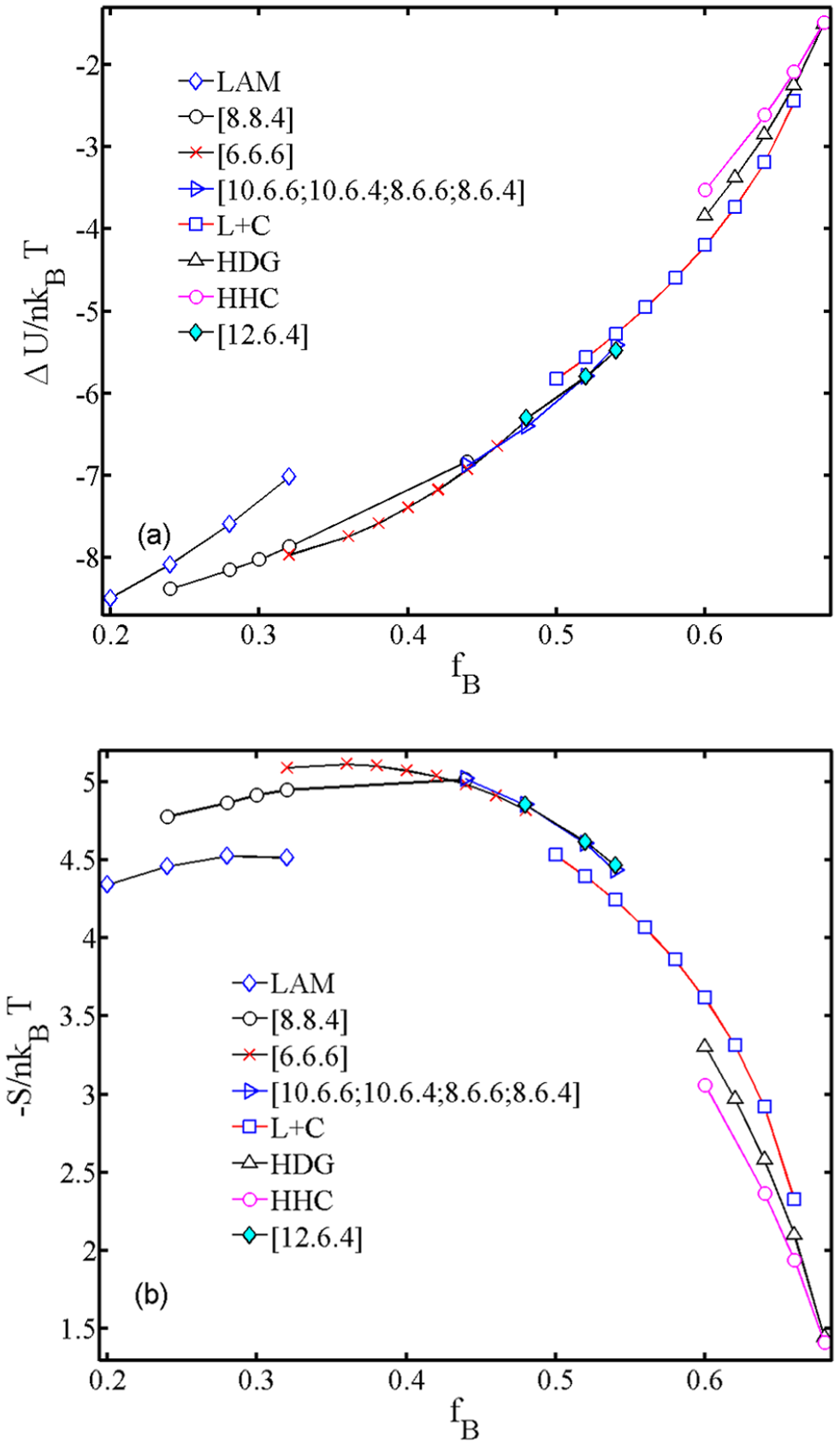}
	  \caption{(a) Internal energy of $\Delta U/nk_B T$ and
	  (b) entropic energy of $-S/nk_B T$ of various structures as a
  function of $f_B$ on the phase path of $f_A=f_C$.}
\label{fig:energysplit1x1}
\end{figure}
After splitting the SCFT energy functional into internal energy
and entropic energy as expressed in Eqn.\,\eqref{eqnenergysplit}
(see Fig.\,\ref{fig:energysplit1x1}), the stability of 3D hierarchical
structures HHC and HDG along the phase path can be understood more readily.
When the ratio of $x=f_B/f_A$ ($f_A=f_C$) is large enough, the
asymmetric ABC star triblock copolymers will
exhibit similar phase behavior of asymmetric diblock copolymer.
The longest B arm plays an equal role as the long block in an asymmetric diblock
copolymer, whereas the short A arm, together with C arm, are just
like the short block in the diblock copolymer.
In asymmetric diblock copolymers, the systems
tend to curve the interface towards the minority domain.
It requires the minority block to stretch, and the cost is more
than compensated for by relaxation of the longer blocks,
which increases the internal energies\,\cite{matsen2002standard}.
Besides the asymmetry of blocks, the segregation power between
arms A and C, $\chi_{AC}N$, is sufficiently large to separate the A-, and C-domains,
leading to the formation of hierarchical structures.
In particular, for the asymmetric ABC star copolymers, the
blocks A and C have shorter chains than the short block in AB
diblock when the curve interface can be formed.
The A and C arms can freely relax in their packing frustration
domains which greatly reduces the entropic energy of the system.
Although the internal energies of 3D hierarchical structures
are higher than that of L+C structure, their entropic energies are much lower than that of the
2D cylinders-in-lamella phase.
As a consequence, the combination of the two competing energies
makes 3D hierarchical structures stable rather than L+C when $f_B\geq 0.63$.
At higher asymmetries, block copolymers prefer to form structures with a higher
interfacial curvature\,\cite{matsen1996origins}.
Therefore HHC phase with larger spontaneous curvature than that
of HDG phase, has lower free energy when $f_B\geq 0.67$.

\subsection{Triangular Phase Diagram}

\begin{figure}[!htbp]
  \centering
	  \includegraphics[scale=0.5]{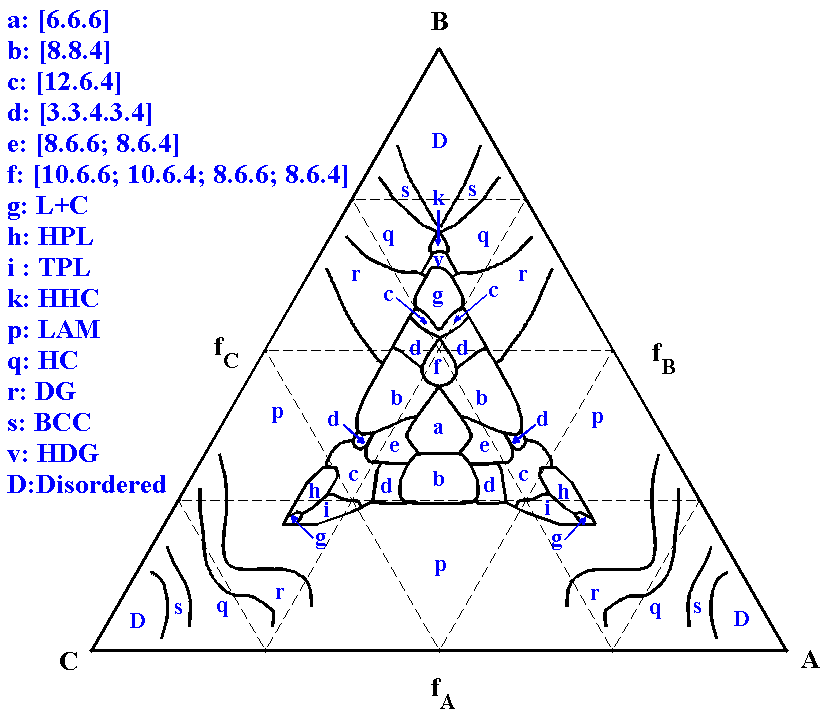}
  \caption{The triangular phase diagram of ABC star triblock
  copolymers with equal statistical segment lengths for each
  block and with asymmetric interaction parameters of
  $\chi_{AC}N=50.0$, $\chi_{AB}N=\chi_{BC}N=30.0$.
The ordered structures emerging in diblock
copolymers, i.e., no core-shell cylinders, spheres, gyroid, two-color lamella, are not included in our calculations.
That is, near the boundary of the triangular phase diagram,
simulations are not carried out in our study.
  }
  \label{fig:phasediagram}
\end{figure}

The phase transition sequence for systematically varying volume
fractions can be obtained by repeating the free energy comparison among the candidate structures.
The results of the phase transition sequences can be summarized in terms
of phase diagrams.
For the case of asymmetric interaction parameters of
$\chi_{AC}N>\chi_{AB}N=\chi_{BC}N$, the triangular phase
diagram is mirror symmetric with the axes of $f_A=f_C$.
Therefore one-half of the whole triangular phase diagram should be calculated.
The triangular phase diagram obtained by our SCFT simulations is presented in
Fig.\,\ref{fig:phasediagram}. Sixteen
structures, including 2D, 3D ordered phases and disordered phase
(D), are predicted to be stable in the phase
diagram. Besides the [6.6.6], [8.8.4],
[8.6.6; 8.6.4], [3.3.4.3.4], [10.6.6;10.6.4;8.6.6;8.6.4],
[12.6.4], L+C, TPL, HHC, HDG and LAM phases
discussed above, four more ordered structures, HPL, core-shell structures of
HC, DG, and BCC, and disordered phase are included in the triangular phase diagram.
The regions of stability of the different phases are obtained by
comparing the free energy of these candidate structures.
The phase boundaries are determined by calculating the cross over
point of the free energies of the two neighbouring phases.
The most significant feature of the triangular phase diagram is
the rich phase behavior with a large number of stable ordered phases.
It should be noted that the ordered structures emerging in diblock
copolymers, i.e., no core-shell cylinders, spheres, gyroid, two-color lamella, are not included in our calculations.
That is, near the boundary of the triangular phase diagram,
simulations are not carried out in our study.

When the chain lengths of three arms are close to each other, junction points are
aligned on a straight line, and hence the system tends to form
polygonal tilings to get smaller internal energy.
As a consequence, the central region of the phase map is dominated by the 2D tiling patterns.
The coordination number of each domain is proportional to the
composition of corresponding block. For example, as discussed above,
along the phase path of $f_A:f_B:f_C$ = $1.0:1.0:x$, the coordination
number of C-domains changes from $4$, to $6$, to $8$, then to
$10$, and finally to $12$ with increasing of $x$ from $0.435$ to $2.76$.
Besides the cylindrical structures observed by experiments,
two tilings of [8.6.6; 8.6.4] and [10.6.6;10.6.4;8.6.6;8.6.4]
are predicted as the stable phases in our theoretical calculations.
The stability has been analyzed in the above context.
It is very different from the phase behavior of ABC linear triblock copolymer melt
in which the lamellar phase occupies the large central region\,\cite{tyler2007scft}.
The reason is attributed to the topology of the star copolymer chain
for which the junction points tend to be aligned on a straight
line when the chain lengths of three arms are comparable.
As the compositions of star copolymer become asymmetric, more 2D and 3D phases,
including LAM, HC, DG, BCC, HPL, TPL and cylinders-in-lamella of L+C, appear and
surround these 2D tilings in the triangular phase diagram.
Among these phases, LAM phase has the following phase transitions.
Near the AB edge, CACB layers can be formed
whereas ABAC layers formed near the BC edge, and BABC layers formed near the AC edge.

Consider the structural evolution that starts from the central 2D tilings region,
toward the A- and C-rich corners of the triangular phase diagram.
The phase transition sequence is the cylinders-in-lamella
structures (including HPL and TPL), L+C, LAM, DG, HC, BCC
and disordered phase.
Near the A-rich corner, in the cylinders-in-lamellar structure,
the C arms form internal cylinders in the B perforated lamellar, then
blocks B and C form lamellar together with A-layers.
The packed manners of the cylinders determine the morphologies of HPL
and TPL, as shown in Fig.\,\ref{fig:phases}.
The 2D hierarchical pattern, L+C, where arms B and C
form alternating cylinders in the lamellar-based phase, has a
stable region between cylinders-in-lamellar structures and LAM phase.
The prediction of stability region of L+C phase is generally consistent with the
recent experiment in which Nunns et al. observed the phase in
$\mathrm{I_{0.20}S_{0.14}F_{0.66}}$ star copolymer system\,\cite{nunns2013synthesis}.
Because of the smallness of $\chi_{BC}$, a further
segregation between the minority components B and C (A and B) will not
occur, which implies that the 3D hierarchical structures will not
be global stable in this region. Instead, the LAM phase dominates this area.
Owing to $\chi_{BC}=\chi_{AB}$,
there is similar phase behavior near the C-rich corner when
exchanging the position of arms A and C.

The BCC, HC, and DG phases form continuous areas across the A- and
C-rich corners of the triangular phase diagram, where A or C block
is the largest arm.
The continuous areas formed by these phases in the
triangular phase diagram reflect a continuous evolution in the
compositions of the core and shell blocks. For example, consider
the evolution of structure along a path within the BCC phase
in the A-rich corner, starting from the AB edge, where the
structure contains B spheres in an A matrix, to the AC edge.
As the C arm increases its length,
a spherical ``shell'' of C grows in the middle of each B-sphere,
while the composition of the surrounding shell of B shrinks,
until a structure of C spheres in an A matrix is obtained at the
AC edge. An analogous change in the volume fractions of the
``shell'' and ``core'' components occurs in the HC and DG
phases in both the A- and C-rich corners.

The phase behavior in the B-rich corner is more complicated.
Owing to the largest value of $\chi_{AC} N$,
the minority components A and C tend to a further segregation
near the B-rich corner,
which leads to the formation of the hierarchical structures.
From the cylindrical regions to B-rich corner, the phase sequence of
hierarchical morphologies is L+C $\rightarrow$ HDG $\rightarrow$ HHC.
Near the $f_A=f_C$ line, the stable regions of HC and DG are separated by hierarchical structures of HHC, HDG,
L+C and three Archimedean tilings of [12.6.4], [3.3.4.3.4], [8.8.4].
The stable region of BCC phase in the B-rich corner is continuous above that of HHC structure.
Consider the structural evolution along a path that starts from the AB edge,
where the length of arms satisfies the relationship
$f_B>f_A>f_C$, toward the $f_A = f_C$ isopleth.
Along this path, for example, in the BCC phase, A arms segregate
into spherical domains, surrounded by C-rich pockets shell within the B matrix.
The core-shell HC and DG phases have similar phase behavior with A-core, C-shell and B-matrix.
Similarly, structures, such as DG, HC, and BCC, evaluating along a path that starts from
BC edge toward the $f_A=f_C$ isopleth, are of the C-core, A-shell
and B-matrix patterns.
In addition, the disordered phase emerges in the B-rich corner of
the triangular phase diagram.

In general, the theoretical phase behavior is in agreement with the experimental observations.
However, there are some discrepancies between our theoretical
computationus and experimental observations.
Some structures observed by experiments in ISP star triblock
copolymers, such as L+S phase\,\cite{takano2007composition},
Zinc-blende type structure\,\cite{matsushita2010jewelry} are not obtained in our simulations.
The stable regions of some structures obtained by experiments are not exactly
located in the predicted phase regions of our phase diagram.
Our theoretical phase diagram predicted [8.6.6; 8.6.4] and
[10.6.6;10.6.4;8.6.6;8.6.4] as stable phases which have been not
observed in the experiments.
There are three main possible reasons for these discrepancies.
The first one is that in the experiments many star triblock
copolymers are obtained by blending two kinds of copolymers, or
adding additional homopolymers. For example, the Zinc-blende type structure was observed in the
$\mathrm{I_{1.0}S_{2.3}P_{0.8}}$ system which was realized by blending the
S homopolymer to $\mathrm{I_{1.0}S_{1.8}P_{0.8}}$.
The homopolymer S can definitely affect the phase stabilities.
The second one is that the predicted phases in the theoretical calculation, such as the
[8.6.6;8.6.4] and [10.6.6;10.6.4;8.6.6;8.6.4], may be metastable
in the recent experimental systems.
The third one is that there are many differences between
theoretical systems and experimental conditions, such as the
different interactions, different monomer sizes.
In our calculations, we neglect the differences between $\chi_{AB}N$ and $\chi_{BC}N$.
The difference is small, however, it might influence the self-assembling behavior.
At the same time, the values of interaction may be different from
the experimental systems.

\section{Conclusions}
\label{sec:concl}

In this work, we have investigated the phase behavior of ABC star triblock copolymers with
asymmetric interaction parameters using the SCFT.
Based on the previous work of screening initialization
strategies\,\cite{xu2013strategy}, we can obtain a large number of
ordered phases in studying the complex polymer systems.
Then we used a fourth-order pseudospectral method, combined with
the Anderson mixing algorithm to calculate the free energy of the observed phases.
Motivated by previously experimental studies, the Flory-Huggins
interaction parameters of $\chi_{BC}N=\chi_{AB}N=30.0$,
$\chi_{AC}N=50.0$ are used to model the experimental systems of
ISP and ISF star copolymers in bulk.
To extend the scope of theoretical study,
a large number of 2D and 3D ordered structures have been involved in our calculations.
In order to shed light on the phase behavior of the ABC star
triblock copolymers, we first determined the phase stability along the phase path $f_A:f_B:f_C=1.0:1.0:x$.
Our results agree with those of ISP
star triblock copolymers for cylindrical phases well.
Then we calculated the phase regions along the phase path $f_A:f_B:f_C=1.0:x:1.0$.
On this phase path, besides the cylindrical structures,
we emphatically analyzed the stability of hierarchical structures of L+C, HDG, HHC.
The phase stability has been analyzed by splitting the SCFT
energy functional into internal and entropic parts in detail.
Finally we constructed a very complicated triangular phase diagram with these candidate structures.
Owing to the case of interactions $\chi_{BC}N=\chi_{AB}N$, the
phase diagram has only one mirror symmetric axis $f_A=f_C$.
Fifteen ordered structures and the inhomogeneous phase constitute the triangular phase diagram.
In general, the phase regions predicted by our SCFT calculations
are consistent with previous theoretical studies of either SCFT
calculations or MC, DPD simulations which mainly focus on the equal interaction systems.
However, in our calculations, the interaction parameters are
more closely related to experimental systems, such as ISP and ISF
star triblock copolymers , i.e., the asymmetric interaction parameters $\chi_{AC}N>\chi_{BC}N=\chi_{AB}N$.
It has been found that the asymmetry of the interaction parameters plays
a profound role in the complex phase formation.
Furthermore, our predicted phase diagram involves more phases, especially 3D structures,
and presents more comprehensive phase behavior.
Our calculations and analysis can be helpful to understand the
self-assembling mechanism of complex structures.
The resulting phase behavior extends the theoretical study to the
asymmetrically interacting ABC star triblock copolymers.
And the presented phase diagram will be a useful guide for further study of
ABC star triblock copolymers.





\begin{appendix}
\section{Appendix: The Implementation of Anderson Mixing Method}

The $k$ iteration in the Anderson mixing method
begins with the evaluation of new fields from Eqns.\,\eqref{eqscftwa}-\eqref{eqscftincompressibility}
\begin{align}
	\eta^{(k)}(\mathbf{r}) &= \frac{\sum_{\alpha=A,B,C} w^{(k)}_{\alpha}(\mathbf{r}) X_\alpha - 2N\chi_{AB}\chi_{BC}\chi_{AC}}{\sum_\alpha X_\alpha},
	\\
	\bar{w}_A^{(k)}(\mathbf{r}) &=
	\chi_{AB}N\phi^{(k)}_B(\mathbf{r}) + \chi_{AC}N\phi^{(k)}_C(\mathbf{r}) + \eta(\mathbf{r}),
	\\
	\bar{w}_B^{(k)}(\mathbf{r}) &=
	\chi_{AB}N\phi^{(k)}_A(\mathbf{r}) +
	\chi_{BC}N\phi^{(k)}_C(\mathbf{r}) + \eta(\mathbf{r}),
    \\
	\bar{w}_C^{(k)}(\mathbf{r}) &=
	\chi_{AC}N\phi^{(k)}_A(\mathbf{r}) +
	\chi_{BC}N\phi^{(k)}_B(\mathbf{r}) + \eta(\mathbf{r}).
\end{align}
In the above expressions, $w^{(k)}_{\alpha}$, $\alpha = A,B,C$,
are the old fields,
$X_A=\chi_{BC}(\chi_{AB}+\chi_{AC}-\chi_{BC})$,
$X_B=\chi_{AC}(\chi_{BC}+\chi_{AB}-\chi_{AC})$,
$X_C=\chi_{AB}(\chi_{AC}+\chi_{BC}-\chi_{AB})$.
Next we evaluate the deviation,
\begin{align}
d^{(k)} = \bar{w}^{(k)}(\mathbf{r}) - w^{(k)}(\mathbf{r}),
\label{}
\end{align}
where $d^{(k)}=(d^{(k)}_{A}(\mathbf{r}),
d^{(k)}_{B}(\mathbf{r}), d^{(k)}_{C}(\mathbf{r}))^{T}$,
$\bar{w}^{(k)}(\mathbf{r})=(\bar{w}^{(k)}_{A}(\mathbf{r}),
\bar{w}^{(k)}_{B}(\mathbf{r}), \bar{w}^{(k)}_{C}(\mathbf{r}))^{T}$,
$w^{(k)}(\mathbf{r})=(w^{(k)}_{A}(\mathbf{r}),$

\noindent $ w^{(k)}_{B}(\mathbf{r}), w^{(k)}_{C}(\mathbf{r}))^{T}$.
From the deviation we can specify an error tolerance through the inner product
\begin{align}
	(g(\mathbf{r}), f(\mathbf{r}))= \frac{1}{V}\int\,d\mathbf{r}
	g(\mathbf{r})f(\mathbf{r}),
	\label{}
\end{align}
where $g(\mathbf{r})$ and $f(\mathbf{r})$ are arbitrary functions.
The error tolerance is defined by
\begin{align}
	\varepsilon_1 =
	\left[\frac{\sum_{\alpha=A,B,C}(d^{(k)}_{\alpha}(\mathbf{r}), d^{(k)}_{\alpha}(\mathbf{r}))}
	{\sum_{\alpha=A,B,C}(w^{(k)}_{\alpha}(\mathbf{r}), w^{(k)}_{\alpha}(\mathbf{r}))}\right]^{1/2}
	\label{}
\end{align}
as a measure of the numerical inaccuracy in the field
Eqns.\,\eqref{eqscftwa}-\eqref{eqscftwc}.

The simple mixing method is performed until a certain
tolerance is reached where a morphology has begun to develop.
From our experience, $\varepsilon_1 < 10^{-2}$, is sufficient in
most cases. We then switch to the Anderson mixing procedure by the
previous $n$ steps to update fields. We assemble the symmetric
matrix in this way
\begin{align}
	U_{ij}^{(k)} = (d^{(k)}-d^{(k-i)}, ~d^{(k)}-d^{(k-j)}),
	\label{eqandersonmatrix}
\end{align}
for $i, j = 1,2,\dots,n$, and vector
\begin{align}
	V_{i}^{(k)} = (d^{(k)}-d^{(k-i)}, ~d^{(k)}) =
	(d^{(k)},~d^{(k)}) - (d^{(k-i)}, ~d^{(k)}).
	\label{eqandersonvector}
\end{align}
From these, we calculate the coefficients
\begin{align}
	C_i = \sum_j^n (U^{-1})_{ij}V_j,
	\label{}
\end{align}
and combine the previous histories as
\begin{align}
	T_{\alpha}^{(k)} &= w_{\alpha}^{(k)}+\sum_{i=1}^{n} C_i
	(w_{\alpha}^{(k-i)}- w_{\alpha}^{(k)}),
	\\
	D_{\alpha}^{(k)} &= d_{\alpha}^{(k)}+\sum_{i=1}^{n} C_i
	(d_{\alpha}^{(k-i)}- d_{\alpha}^{(k)}).
	\label{}
\end{align}
Finally, the new fields are obtained from
\begin{align}
	w_{\alpha, j}^{(k+1)} = T_{\alpha, j}^{(k)} + \lambda
	D_{\alpha, j}^{(k)},
	\label{}
\end{align}
where $0< \lambda = 1.0-0.9^k < 1$\,\cite{matsen2009fast}.

In our implementation, the used previous steps are usually much
less than the number of basis functions or grid points. Therefore
assembling the $n$-order linear system spends more computation
time than solving this system. To save computational amount,
we decompose Eqn.\,\eqref{eqandersonmatrix} into
\begin{align}
	U_{ij}^{(k)}
	= (d^{(k)}, ~d^{(k)}) - (d^{(k-i)}, ~d^{(k)})
	- (d^{(k)}, ~d^{(k-j)}) + (d^{(k-i)}, ~d^{(k-j)}).
	\label{eqandersondecompose}
\end{align}
In $k$ iteration, only the terms related to $d^{(k)}$ in the right term of
Eqn.\,\eqref{eqandersondecompose} are required to calculate, but
the last terms $(d^{(k-i)}, ~d^{(k-j)})$, $i,j=1,\dots,n$, should
be not computed repeatedly which will save the main computational cost in assembling matrix $U$.
In practical calculations, we store the following inner product matrix
\begin{equation}
S=\left(
  \begin{array}{cccc}
	(d^{(k)}, d^{(k)}) & (d^{(k)}, d^{(k-1)})  & \cdots & (d^{(k)}, d^{(k-n)}) \\
	(d^{(k-1)}, d^{(k)}) & (d^{(k-1)}, d^{(k-1)})  & \cdots & (d^{(k-1)}, d^{(k-n)}) \\
    \vdots & \vdots & \ddots & 0 \\
	(d^{(k-n)}, d^{(k)}) & (d^{(k-n)}, d^{(k-1)})  & \cdots & (d^{(k-n)}, d^{(k-n)})
  \end{array}
\right).
\end{equation}
Then we can assemble the matrix $U$ and vector $V$ according to the
expressions of Eqns.\,\eqref{eqandersondecompose} and \eqref{eqandersonvector} using the elements of matrix $S$. Note that the inner product
matrix is symmetric, therefore only a row (or a column, equivalently) of $S$ is
required to update in each iteration.

\end{appendix}



\end{document}